\documentclass[aps,prl,twocolumn,superscriptaddress,nofootinbib,showpacs=false]{revtex4-2}
\usepackage{amsmath,amssymb,bm}
\usepackage{graphicx}
\usepackage[colorlinks=true,linkcolor=blue,citecolor=blue,urlcolor=blue]{hyperref}

\begin{document}

\title{Comment on ``Test of the Gravitational Force Law on Cosmological Scales Using the Kinematic Sunyaev--Zeldovich Effect''}

\author{Alessandro Melchiorri}
\affiliation{Department of Physics, Sapienza University of Rome, Piazzale Aldo Moro 5, 00185 Rome, Italy}

\date{\today}

\begin{abstract}
Gallardo \textit{et al.} [Phys. Rev. Lett. \textbf{136}, 151002 (2026)] constrain the index of the force law $g\propto r^{-n}$ with pairwise kinematic Sunyaev--Zeldovich measurements, $n=2.1\pm0.3$, and exclude the deep-MOND value $n=1$ at $3.3\sigma$. The exclusion does not follow from their calculation. Their templates for $n\neq2$ [their Eq.~(8)] raise the \emph{pair-averaged} Newtonian acceleration to the power $n/2$, whereas the pair average of the pointwise law [their Eq.~(30) for deep MOND] is the average of the transformed field; the step that identifies the two attributes to the three-point function a correction that is of second order and not small. In a Gaussian linear field with linearly biased tracers, Gaussian integration by parts shows that the correctly averaged template is \emph{exactly} proportional to the Newtonian one, with a constant that the analysis absorbs into the mean optical depth. Gaussian realizations with peak-selected tracers find the same proportionality to within 2\% (0.5\% for the fiducial selection), and refitting the data of the Letter with the resulting template gives $\chi^2=20.1$--$20.2$, against $20.1$ for the Newtonian template and $33.8$ for Eq.~(8): in every controlled case examined the data do not discriminate between $n=1$ and $n=2$, and the $3.3\sigma$ exclusion, with the $10\sigma$ forecast that rescales it, rests on the order in which the average and the power were taken. What a self-consistent MOND cosmology predicts for this observable is a separate question, which the Letter does not address.
\end{abstract}

\maketitle

\textit{The claim.}---Gallardo \textit{et al.}~\cite{Gallardo2026} measure the mean pairwise velocity of SDSS BOSS galaxies through the kSZ effect in Atacama Cosmology Telescope maps over 30--230~Mpc, fit a family of velocity templates labelled by the index $n$ of a force law $g\propto r^{-n}$, obtain $n=2.1\pm0.3$, and exclude the deep-MOND value $n=1$ at $3.3\sigma$. The templates are built from the observed galaxy correlation function $\xi_{gg}$. With the public pipeline and data products~\cite{pipeline} I reproduce these numbers within rounding: $n=2.14\pm0.35$ from the 15 bins between 25 and 225~Mpc, with $\chi^2=20.10$ for the $n=2$ template and $\chi^2=33.79$ for the $n=1$ template ($\Delta\chi^2=12.5$ at $n=1$ along the profile). The measurement is not in question. The construction of the $n\neq2$ templates is, and with it the constraints on the force law that the Letter derives from that family of templates.

\textit{Where the error is.}---The Letter parametrizes the force law pointwise as $g=g_0\,[G_b]^{n/2}$ [its Eq.~(7)], where $G_b$ is the dimensionless baryonic (Newtonian) acceleration, so that for a point source $g\propto r^{-n}$. It then obtains the mean pairwise acceleration [its Eq.~(8)] by raising the \emph{pair-averaged} Newtonian acceleration, the linear-theory expression in terms of $\xi_{gg}$, to the power $n/2$:
\begin{equation}
T_n(r)\;\propto\;\big|\langle X_N\rangle_{12}(r)\big|^{n/2},\qquad X_N\equiv(\mathbf g_2-\mathbf g_1)\cdot\mathbf e_{12}\,,
\label{eq:wrong}
\end{equation}
where $\mathbf g_a$ is the Newtonian acceleration at tracer $a$, $\langle\cdot\rangle_{12}$ denotes the average over tracer pairs at separation $r$, and the sign of the infall ($\langle X_N\rangle_{12}<0$) is carried by the fitted amplitude, as in the Letter's pipeline. A force law, however, acts at each point. What the Letter's Eq.~(7) predicts for the pairwise statistic is the pair average of the pointwise transformed acceleration,
\begin{equation}
\langle X_n\rangle_{12}=\big\langle\big[\mathbf F(\mathbf g_2)-\mathbf F(\mathbf g_1)\big]\cdot\mathbf e_{12}\big\rangle_{12},\quad
\mathbf F(\mathbf g)\propto \mathbf g\,|\mathbf g|^{\,n/2-1},
\label{eq:right}
\end{equation}
which for deep MOND is $\mathbf F(\mathbf g)=\sqrt{a_0}\,\mathbf g/\sqrt{|\mathbf g|}$, the Letter's Eq.~(30). Linearity guarantees that Eqs.~(\ref{eq:wrong}) and (\ref{eq:right}) coincide, which is the case $n=2$; for a nonlinear map the average of the function and the function of the average differ in general. Equation~(\ref{eq:wrong}) is the quantity that was fit to the data.

The Letter makes the identification explicitly in its Appendix (``An example: MOND''). From Eq.~(30) it writes $\langle g_M\rangle=\langle\sqrt{a_0 g_b}\rangle$, then ``Taylor expands'' this as $\sqrt{a_0\langle g_b\rangle}+\mathcal O(\zeta)$, attributes the correction to the three-point function $\zeta$, and drops it ``as above'', that is, as in the $\Lambda$CDM derivation, where $\zeta$ vanishes at linear order. This is the error. The expansion of $\langle\sqrt{g}\rangle$ about $\sqrt{\langle g\rangle}$ is an expansion in $(g-\langle g\rangle)/\langle g\rangle$, and its leading correction, $-\langle(g-\langle g\rangle)^2\rangle/8\langle g\rangle^{3/2}$, is a two-point quantity: it does not vanish for a Gaussian field with $\zeta\equiv0$, and it has nothing to do with the three-point function. In the $\Lambda$CDM derivation the neglect of $\zeta$ is legitimate because the map is linear and $\zeta$ is the first correction that appears; for a square root the variance appears first, and the argument does not transfer. Worse, the expansion is taken about the wrong point. The acceleration at a halo is dominated by its whole environment, while $\langle g_b\rangle$ in Eq.~(8) is the small mean shift induced by a partner at 30--230~Mpc. In the realizations described below, the ratio of the rms acceleration at a tracer to this conditional mean is 5.5 at 25~Mpc, 10 at 55~Mpc, 23 at 105~Mpc and 90 at 225~Mpc, so the expansion is not controlled at any order.

\textit{The correct average.}---The pair average of the Letter's pointwise law, Eq.~(\ref{eq:right}), can be computed exactly in a controlled model built from the ingredients of the Letter's template: a Gaussian linear density field $\delta$ (smoothed on the tracer scale), whose Newtonian acceleration $\mathbf g$ is a linear functional of $\delta$, and tracers of linear bias $b$, which enter the pair statistic through the weight $W_{12}=(1+b\delta_1)(1+b\delta_2)$, $\delta_a\equiv\delta(\mathbf x_a)$. The Letter does not state its assumptions in this form; this is the simplest model in which the prescription of Eq.~(8) can be checked exactly, and the numerical test below goes beyond it. The pair average is
\begin{equation}
\langle Y\rangle_{12}=\frac{\langle W_{12}\,Y\rangle}{\langle W_{12}\rangle}\,.
\label{eq:pairavg}
\end{equation}
All of $(\delta_1,\delta_2,\mathbf g_1,\mathbf g_2)$ are jointly Gaussian with zero mean. Let $F_i(\mathbf g)=g_i|\mathbf g|^{p}$ with $p=n/2-1$ ($p=-1/2$ for deep MOND). For jointly Gaussian variables, Gaussian integration by parts (Stein's identity~\cite{Stein1981}) gives
\begin{equation}
\langle\delta_a\,F_i(\mathbf g_2)\rangle=\langle\delta_a\,g_{2,j}\rangle\Big\langle\frac{\partial F_i}{\partial g_j}\Big\rangle
=\kappa_p\,\langle\delta_a\,g_{2,i}\rangle,
\label{eq:stein}
\end{equation}
with, by isotropy,
\begin{equation}
\Big\langle\frac{\partial F_i}{\partial g_j}\Big\rangle=\delta_{ij}\,\kappa_p,\qquad
\kappa_p=\Big(1+\frac p3\Big)\big\langle|\mathbf g|^{p}\big\rangle,
\label{eq:kappa}
\end{equation}
that is, $\kappa_p=\frac{n+4}{6}\langle|\mathbf g|^{\,n/2-1}\rangle$, which for deep MOND is $\kappa=\tfrac56\langle|\mathbf g|^{-1/2}\rangle$. The remaining terms of $\langle W_{12}F_i(\mathbf g_2)\rangle$ vanish identically: $\langle F_i(\mathbf g_2)\rangle=0$ by parity; $\langle\delta_2\,g_{2,i}\rangle=0$ because a scalar and a vector at the same point are uncorrelated in an isotropic field; and, applying Eq.~(\ref{eq:stein}) twice, $\langle\delta_1\delta_2F_i(\mathbf g_2)\rangle=\langle\delta_1g_{2,j}\rangle\langle\delta_2g_{2,k}\rangle\langle\partial_j\partial_kF_i\rangle=0$. Hence
\begin{equation}
\langle W_{12}\,F_i(\mathbf g_2)\rangle=\kappa_p\,b\,\langle\delta_1\,g_{2,i}\rangle=\kappa_p\,\langle W_{12}\,g_{2,i}\rangle,
\end{equation}
since the other contributions to $\langle W_{12}\,g_{2,i}\rangle$ vanish for the same reasons. The same holds at point 1. Therefore, exactly and at every separation,
\begin{equation}
\langle X_n\rangle_{12}(r)=C_n\,\langle X_N\rangle_{12}(r),\qquad C_n\propto\frac{n+4}{6}\big\langle|\mathbf g|^{\,n/2-1}\big\rangle .
\label{eq:prop}
\end{equation}
No expansion in the conditional mean shift and no assumption about the conditional distribution are involved. In this model the pair average of the Letter's pointwise law, Eq.~(\ref{eq:right}), has the Newtonian radial shape for all indices considered here (the moments in Eq.~(\ref{eq:kappa}) and in the second-derivative term exist for $n>-2$); the entire $n$ dependence of the shape of Eq.~(\ref{eq:wrong}) is the artifact of taking the power after the average. Outside this model, exact proportionality is not guaranteed, and the question becomes numerical.

\begin{figure}[t]
\includegraphics[width=\columnwidth]{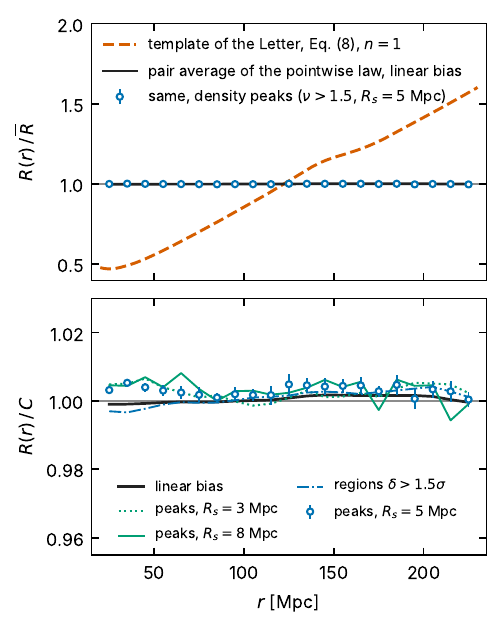}
\caption{Ratio $R(r)=\langle X_M\rangle_{12}/\langle X_N\rangle_{12}$ of the pair-averaged deep-MOND to Newtonian acceleration in Gaussian realizations. Top: $R$ normalized to its mean over 25--225~Mpc, for linear-bias weights (black; the model in which Eq.~(\ref{eq:prop}) is exact) and for tracers at density peaks above $1.5\sigma$ (points; error bars are errors on the mean over six realizations). Dashed: the ratio implied by the Letter's template of Eq.~(8) for $n=1$, the shape that was fit to the data, in the same normalization. Bottom: $R$ divided by the constant $C=\tfrac56\sqrt{a_0}\langle|\mathbf g|^{-1/2}\rangle$ of Eq.~(\ref{eq:prop}), evaluated at the tracers, for the same selections and for peaks at $R_s=3$ and 8~Mpc and thresholded regions.}
\label{fig:ratio}
\end{figure}

\textit{Numerical check.}---As a controlled test bed I use Gaussian realizations of the Planck-2018 linear density field at $z=0.55$ (the proportionality in Eq.~(\ref{eq:prop}) does not depend on the spectrum, only the constant $C_n$ does; the Letter's own template uses the observed $\xi_{gg}$), on periodic boxes of 1280~Mpc with $320^3$ cells (and 960~Mpc with 3-Mpc cells), smoothed with Gaussian filters of $R_s=3$, 5 and 8~Mpc. The physical acceleration at the tracers has rms $1.7\times10^{-12}\,\mathrm{m\,s^{-2}}$, well inside the deep-MOND regime. Tracers are selected as peaks of the smoothed density above $\nu=1$, 1.5 and 2 (number densities from $10^{-5}$ to $10^{-4}\,\mathrm{Mpc^{-3}}$), a nonlinear selection outside the linear-bias model; the linear-bias weights $1+2\delta$ and thresholded regions $\delta>1.5\sigma$ are run alongside. The deep-MOND map is applied pointwise to the Newtonian acceleration of each tracer, and only then are pair differences and the conditional average taken; the pair average is evaluated exactly over all pairs on the grid (over $10^7$ peak pairs per realization, in 10-Mpc shells from 20 to 230~Mpc). Six realizations of the 1280-Mpc box and two of the 960-Mpc box are used; the uncertainties quoted are errors on the mean. Figure~\ref{fig:ratio} shows the ratio $R(r)=\langle X_M\rangle_{12}/\langle X_N\rangle_{12}$. For the linear-bias weights it is constant to 0.3\% peak to peak over 25--225~Mpc in the large box (0.8\% in the small one), with mean equal to $\tfrac56\sqrt{a_0}\langle|\mathbf g|^{-1/2}\rangle$ to $0.04\pm0.12\%$: Eq.~(\ref{eq:prop}) is verified. For the peak-selected tracers, to which the derivation does not apply, $R$ is constant to 0.5--0.7\% at $R_s=5$~Mpc and 1.4--1.6\% at $R_s=8$~Mpc in the large box, and to 0.6--0.9\% ($R_s=3$~Mpc) and 1.1--2.0\% ($R_s=5$~Mpc) in the small box, independently of the threshold; its mean agrees with the same constant, evaluated at the tracers, to $0.3\pm0.15\%$ for the fiducial selection ($L=1280$~Mpc, $R_s=5$~Mpc, $\nu>1.5$) and to better than 0.6\% for every selection, a numerical result that the linear-bias derivation does not guarantee. The shape ratio implied by the Letter's Eq.~(8) for $n=1$, $|\langle X_N\rangle_{12}|^{-1/2}$, varies by a factor 3.3 over the same range. The scripts reproducing every number quoted here are provided as ancillary files.

\textit{The consequence.}---The amplitude of the template is a free parameter of the fit, degenerate with the mean optical depth $\bar\tau$. The constant $C_n$ is therefore unobservable. With the exact proportionality~(\ref{eq:prop}) the deep-MOND and Newtonian templates give the same $\chi^2$ by construction. Refitting the data of the Letter with its pipeline and covariance (the template covariance propagated through the factor $R$), using the Newtonian template multiplied by the ratio $R(r)$ measured in the realizations, gives $\chi^2=20.10$--$20.17$ for the linear-bias weights and $\chi^2=20.08$--$20.20$ for every peak selection, smoothing scale and threshold, against $\chi^2=20.10$ for the Newtonian template and $33.79$ for the template of Eq.~(\ref{eq:wrong}) at $n=1$. Including the estimated covariance of the simulated mean ratio in the fit changes the minimum $\chi^2$ by less than $0.004$; the realization-to-realization scatter of the mean curve propagates, by jackknife, to an uncertainty of $0.04$ on the minimum $\chi^2$ for the fiducial selection and of at most $0.07$ in the large box. The $3.3\sigma$ exclusion of $n=1$ comes entirely from the radial shape of Eq.~(\ref{eq:wrong}) at $n=1$, which is not the shape of the pair-averaged law, the Letter's Eq.~(30). In every controlled case examined the data have no discriminating power between $n=1$ and $n=2$, nor, within the Gaussian linear-bias model, between any two values of $n$. The forecast that upcoming surveys will exclude $n=1$ at $10\sigma$ rescales the present $3.3\sigma$ by $\sqrt{N_{\rm gal}}$ and inherits the template; it is void for the same reason. The constraint $n=2.1\pm0.3$ survives as a statement about the exponent of the phenomenological family~(\ref{eq:wrong}); it is not a constraint on the index of the Letter's force law, its Eq.~(7), because for $n\neq2$ the family~(\ref{eq:wrong}) is not the pairwise response to that law.

\textit{Scope.}---The result above is a statement about the Letter's prescription: when the average and the power are taken in the correct order, in the controlled model and in the realizations examined here, the deep-MOND template becomes degenerate with the Newtonian one. It is not a MOND prediction for the pairwise kSZ signal, and the Letter does not contain one either, for three further reasons. (i) The pointwise relation (30) is justified in the Letter by the curl-free property of $\mathbf u=\mu\,\mathbf g_M$ ``for problems with planar symmetry'', on the ground that three sources are considered at a time. The field at a halo is sourced by the whole density field; $\mathbf u=\mathbf g_N+\nabla\times\mathbf h$ in AQUAL~\cite{BM84}, and planar, cylindrical, or spherical symmetry are sufficient conditions for the solenoidal term to vanish, not properties of a generic configuration. Constraints on primordial non-Gaussianity, invoked in the Letter at this point, have no bearing on it. (ii) Equations~(8) and (30) are accelerations. The observed pairwise velocity requires integrating the modified dynamics over the growth history, with initial conditions, and the Letter supplies neither. (iii) The alternative prescription of its Eq.~(32) does not restore the exclusion, since the Letter itself states that it does not correspond to MOND for any $n$. A MOND prediction for this observable requires a theory-specific field equation and a cosmological growth history, neither of which appears in the Letter.

\textit{Conclusion.}---The deep-MOND template of Gallardo \textit{et al.} is the force law applied to the average instead of the average of the force law, and the step that identifies the two, immediately after their Eq.~(30), attributes to the three-point function a correction that is of second order and not small. In a Gaussian linear field with linearly biased tracers, the correctly averaged template is exactly proportional to the Newtonian one for all indices considered here; in realizations with peak-selected tracers it remains so to within 2\%, and refitting the data with it recovers the Newtonian $\chi^2$. In every controlled case examined the data fit $n=1$ and $n=2$ equally well. The $3.3\sigma$ exclusion of deep MOND and the $10\sigma$ forecast rest on the order of operations, and the Letter has not confronted the data with a MOND prediction. The pairwise kSZ effect may yet become a powerful probe of gravity on cosmological scales; establishing that requires a template that follows from a field equation, which remains to be computed.

\end{document}